# Interplay of crystal symmetries and light's topology in high harmonic spectroscopy

Ana García-Cabrera,[*] Roberto Boyero-García, Óscar Zurrón-Cifuentes,
Javier Serrano, Julio San Román, Luis Plaja, and Carlos Hernández-García

*Grupo de Investigación en Aplicaciones del Láser y Fotónica,*
*Departamento de Física Aplicada, Universidad de Salamanca, E-37008, Salamanca, Spain*

(Dated: November 3, 2023)

## Abstract

Structured ultrafast laser beams offer unique opportunities to explore the interplay of the angular momentum of light with matter at the femtosecond scale. Linearly polarized vector beams are paradigmatic examples of structured beams whose topology is characterized by a well-defined Poincaré index. It has been demonstrated that the Poincaré index is a topological invariant during high-order harmonic generation from isotropic targets, such as noble gases. As a result, harmonics are produced as extreme-ultraviolet vector beams, with the same topology as the driver. We demonstrate that this simple conservation rule does not apply to crystalline solids, characterized by their anisotropic non-linear response to the driving excitation. In this context, we identify the topological properties of the harmonic field as unique probes, sensitive to both the microscopic and macroscopic features of the target's complex non-linear response. Our simulations, performed in single-layer graphene but extendable to other solid targets, show that the harmonic field is split into a multi-beam structure whose topology—different from that of the driver—encodes information about laser-driven electronic dynamics. Our work opens the route towards using the topological analysis of the high-order harmonic field as a novel spectroscopic tool to reveal the coupling of light and target symmetries in the non-linear response of matter.

---

[*] anagarciacabrera@usal.es

I. INTRODUCTION

Nonlinear optics stands nowadays as a unique approach not only to up-convert laser radiation to higher frequencies but also to obtain information on the dynamics of laser-driven media. High-harmonic generation (HHG) represents an extraordinary example, capable of producing extreme-ultraviolet (EUV) to soft x-ray coherent radiation, as well as of unveiling electronic dynamics at the attosecond scale [1]. In gases, HHG can be readily understood using a semiclassical point of view [2]. According to it, first, an intense infrared laser field liberates an electronic wavepacket from the atom via tunnel ionization and accelerates it in the continuum. Then, upon reversal of the field amplitude, the electronic wavepacket is redirected to the parent ion, where it recombines emitting high-frequency radiation. The radiated spectrum contains unique information about the ultrafast non-perturbative electron dynamics during the interaction. As a result, high-harmonic spectroscopy has emerged as a main technique to access the ultrafast dynamics of matter subjected to intense laser fields [3–5].

It was not until recently that HHG in crystalline solids was demonstrated, so the full potentialities of these targets are currently being unravelled [6, 7]. For driving beams at grazing incidence, HHG from solids is mediated by electrons detached from the target and, therefore, has a resemblance to its atomic counterpart [8]. However, despite this parallelism, it has been shown that the periodicity of the crystal imprints a diffraction pattern in the electronic wavefunction, giving rise to Talbot revivals, with signatures in the harmonic spectrum [9]. In contrast, when driven at normal incidence, HHG from solids can be interpreted in terms of semi-classical trajectories of electron-hole pairs in the target, excited via tunnelling or Landau-Zener transitions, which subsequently evolve accordingly to the band's energy dispersion [10]. In this case, the harmonic emission takes place upon recombination of the electron-hole pair, following either perfect or imperfect recollisions [11–13]. As in gas targets, high-harmonic spectroscopy of solids has emerged as a fundamental technique giving access to information about intraband currents in bulk solids [14], the Berry curvature [15], many-body dynamics in strongly correlated systems [16], or the ultrafast dynamics of carriers [17, 18], among others.

During the last decade, there has been considerable interest in driving HHG with structured laser beams, in order to obtain coherent short-wavelength radiation with controlled spin (SAM) and/or orbital (OAM) angular momentum. Whereas SAM is connected to the field polarization—characterized by the spin index, $\sigma = -1$ for right (RCP) and $\sigma = +1$ for left (LCP) circularly polarization states— OAM is associated with the beam's azimuthal phase variation [19], and it is characterized by the topological charge, a discrete index that can take infinite integer values. Such laser sources are valuable tools for the ultrafast control of electronic currents at the nanoscale [20, 21].

It is not trivial to convey the angular momentum properties of the driving beam into high-order harmonic radiation. For instance, in the case of gaseous targets, the efficiency of HHG drops drastically when driven by elliptically polarized fields [22]. Nevertheless, it is still possible to produce harmonic radiation with on-demand SAM from atomic targets by using rather sophisticated driver geometries, such as bicircular fields [23] or noncollinear beams [24, 25], among others [26–29]. In contrast, the topological charge of the high-order harmonics driven by linearly polarized single-OAM beams scales linearly with that of the driving field [30–32]. The deep understanding of OAM-SAM conversion in HHG from gaseous targets, which requires a macroscopic description, has inspired the engineering of a wide variety of schemes that allow for the fine spatiotemporal control of the intensity, phase and polarization properties of the high-order harmonics [33–38]. In this context, vector beams are particularly interesting. These beams result from the combination of ravelled SAM and OAM modes. Among them, linear-polarized vector beams (LPVB) present a transversal distribution of linearly polarized states with different tilt angles [39]. This azimuthally-varying orientation confers the beam with a topological character, with a well-defined Poincare index [40]. In this sense, it has been already demonstrated that the up-conversion of LPVBs to high-order harmonics in gases preserves the topology of the driving field [41, 42].

The general scenario of SAM-OAM conversion in HHG changes completely in the case of solid targets, in particular for crystals, where symmetries can introduce anisotropy in their nonlinear optical response [43–48]. The exploration of the interplay of the electromagnetic field topology and the target symmetries in HHG remains barely explored, being limited to the study of OAM conservation in semiconductors [49], to the best of our knowledge. In this sense, we shall see that the analysis of the topological properties of high-order harmonics driven in solids stands as a novel and promising route for high harmonic spectroscopy of condensed matter [50].

In this article, we identify light's topology as a property sensitive to the electronic dynamics in crystals, which establishes the basis of a topological approach to high harmonic spectroscopy. To do so, we explore the up-conversion to high-harmonics of an LPVB driver by single-layer graphene (SLG). The investigation of the coupling of light's topology with crystal symmetries finds a privileged scenario in HHG from two-dimensional crystals driven by LPVB. On the one hand, their atomic-thin thickness excludes the effects of the propagation inside the target. On the other hand, the target presents well-defined symmetries that play a relevant role. As an example, the nonlinear response of SLG is sensitive to the driver's polarization tilt angle, with $\pi/3$ periodicity according to the hexagonal symmetry of the lattice [47]. As a main result, we find that HHG from SLG driven by LPVB produces harmonic beams composed of a central vector beam, that retains the topological characteristics of the driving field, surrounded by a topological cluster encoding specific information about the crystal's anisotropic nonlinear response. Remarkably, therefore, the conservation of the driver's topology in HHG found in isotropic targets [41] is broken in the generation of the topological cluster. We present an analytical model that demonstrates how the topological structure of the harmonic far-field encodes unique information about the crystal's nonlinear response. Indeed, sub-cycle dynamics, such as those arising from interband and intraband transitions, can be also distinguished through the topological structure of the harmonic beam. Therefore, we envisage a novel spectroscopic method that uses the parameters of the far-field topology to unveil details of the nonlinear response of the target. In addition, our work demonstrates that crystalline solid targets presenting a non-linear anisotropic response are extremely interesting playgrounds for the generation of structured short-wavelength radiation with intertwined SAM and OAM properties.

## II. RESULTS

We perform theoretical simulations of HHG in SLG driven by LPVB (see Methods), corresponding to a superposition of two counter-rotating circularly-polarized Laguerre-Gauss beams with opposite topological charges. The driver's transverse profile at the focus can be described as

$$\mathbf{E}(\rho, \varphi, t) = U_0(\rho) e^{-i\omega t} \left[ e^{-i(\ell\varphi - \theta_0)} \mathbf{e}_L + e^{+i(\ell\varphi - \theta_0)} \mathbf{e}_R \right] \quad (1)$$

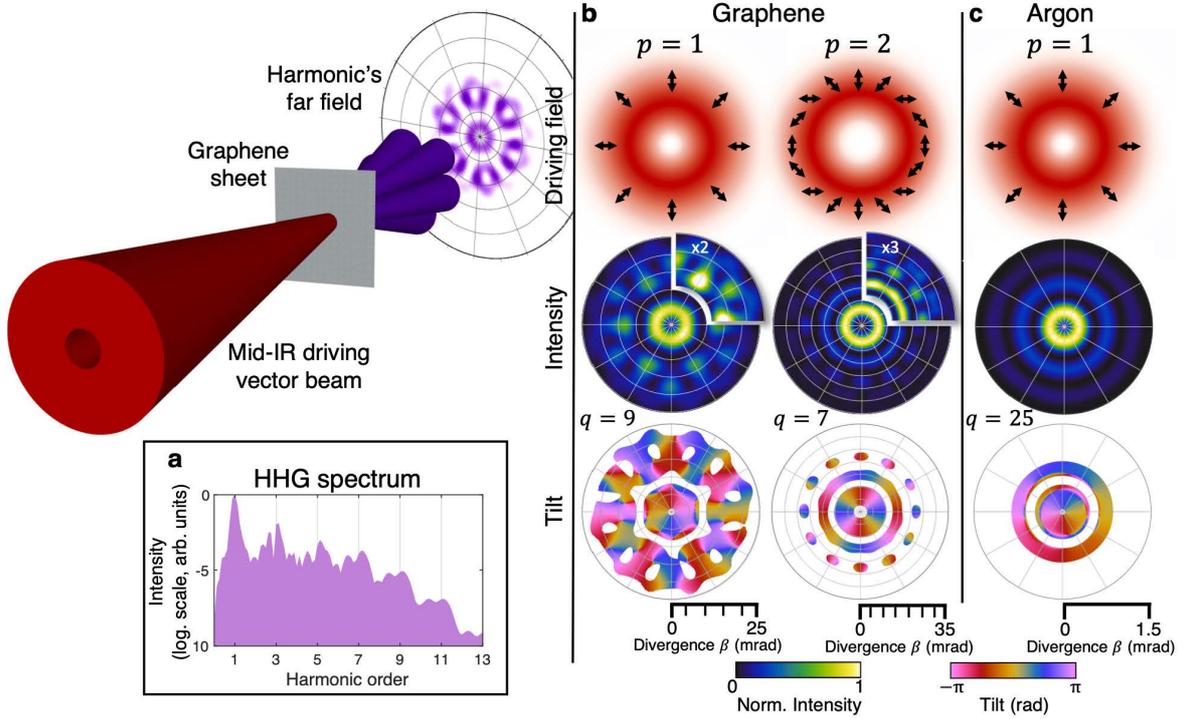

FIG. 1. Scheme of the interaction geometry of HHG in single-layer graphene driven by a LPVB. The driving beam is aimed to a graphene sheet at normal incidence, where HHG results in the emission of high-order harmonics. (a) Spatially-integrated far-field harmonic spectrum emitted by the graphene target when driven by a LPVB with Poincar index $\mathcal{P} = 1$—which corresponds to a radial vector beam. (b) Results of HHG in graphene driven by a LPVB with Poincaré indices $\mathcal{P} = 1$ (left column) and $\mathcal{P} = 2$ (right column). The first row shows the total intensity and polarization profiles of the driving LPVB. The far-field total intensity and polarization tilt—in the regions with an intensity over 10% of the maximum—are shown in the second and third rows, respectively, for the 9th (left) and 7th (right) harmonics. The off-axis far-field intensity is magnified to show the details of the harmonic emission. Panel (c) shows the same features for the 25th harmonic driven in argon by a LPVB with $\mathcal{P} = 1$.

where $\rho$ is the radius, $\varphi$ the azimuth angle, $\ell$ is the absolute value of the OAM charge of the modes composing the LPVB, $\mathbf{e}_{L,R}$ are the left/right polarization vectors, and $\theta_0$ defines the geometry of the beam's polarization. For the particular case of $\ell = 1$, $\theta_0 = 0$ describes a radial vector beam and $\theta_0 = \pm\pi/2$ an azimuthal one. In the following we shall consider a radial LPBV, therefore $\ell = 1$ and $\theta_0 = 0$. Note that, due to the opposite values of OAM and SAM of the composing modes, LPVB are vectorial fields with no net OAM or SAM. However, they are topologically characterized by their Poincaré index, $\mathcal{P}$—the number of complete rotations of the polarization tilt along a closed loop around the axis [40]—which coincides with the OAM charge of the RCP mode, i.e. $\mathcal{P} = \ell$.

The interaction geometry studied in this work is sketched in Fig. 1. We consider an eight-cycle (28 fs full width at half maximum in intensity), 3 $\mu$m wavelength driving pulse, with $\sin^2$ envelope and peak intensity of $5 \times 10^{10}$ W/cm$^2$. The driving field, structured as a LPVB with beam waist 30 $\mu$m, is aimed at normal incidence onto a SLG sheet. Note that tighter focusing conditions, where the paraxial approximation is broken, would induce a non-negligible on-axis longitudinal component [51].

We consider that, after generation at the graphene layer, the high-order harmonics are detected in the far field. Fig. 1a depicts the spatially-integrated far-field harmonic spectrum generated in graphene by a radially polarized vector beam. As for gas targets, the spectrum presents a *plateau* of harmonics, a characteristic signature of the non-perturbative non-linear interaction. For the present driving field, the harmonic *plateau* extends up to the 9th order, followed by a cut-off frequency where harmonic efficiencies decrease at a ratio of approximately one order of magnitude per harmonic interval. The fundamental details of this structure can be understood in semiclassical terms, according to the recollision trajectories of electron-hole pairs excited at the neighborhood of the Dirac points [12].

We present in Fig. 1b results of HHG in graphene driven by LPVB with $\mathcal{P} = 1$ —which corresponds to a radially polarized vector beam (left column)—and $\mathcal{P} = 2$ (right column). The driving field intensity and polarization profiles are shown together with the far-field intensity and linear-polarization tilt angle distributions for two sample harmonics (the 9th harmonic for the $\mathcal{P} = 1$, and the 7th harmonic for the $\mathcal{P} = 2$ driving fields). Results for the rest of high-order harmonics are shown in the Supplementary Information. For the sake of comparison, we show in Fig. 1c the far field of the 25th harmonic obtained in an Ar slab driven by a radially polarized beam ($\mathcal{P} = 1$). For this latter case we have used standard parameters for HHG in Ar (50 $\mu$m-waist, 800 nm wavelength, peak intensity of $1.7 \times 10^{14}$ W/cm$^2$ and same pulse envelope and duration as that used in graphene). Remarkably, while the vector beam character of the driver—intensity profile and $\mathcal{P}$—is translated into to the harmonic emission in Ar, the up-conversion in graphene is much more complex.

To shed light on these results, we have analyzed in detail the two polarization components of the harmonic emission. Fig. 2 shows the far-field intensity and phase distributions of the 9th and 7th harmonics depicted in Fig. 1b for the $\mathcal{P}=1$ and $\mathcal{P}=2$ LPVB drivers, decomposed into their LCP (left column) and RCP (right column) components.

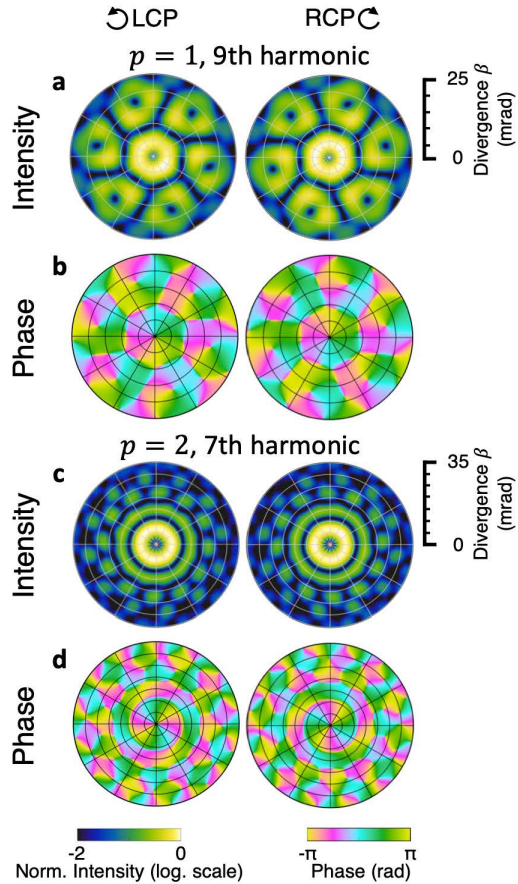

FIG. 2. Far-field intensity (a,c) and phase (b,d) distributions of the 9th and 7th harmonics depicted in Fig. 1b for SLG driven by the $\mathcal{P}=1$ and $\mathcal{P}=2$ LPVB, decomposed into the LCP (left column) and RCP (right column) components.

Very interestingly, the diffraction patterns of the two polarization components are displaced from each other. This demonstrates that SLG's diffraction of the harmonic field is spin-dependent, a consequence of the anisotropic character of its non-linear response. Note also that only at low-divergence angles, both polarization modes fully overlap.

A further analysis of the harmonic far-field characteristics can be drawn exploring the particular OAM composition of each of the polarization modes. To do so, we perform the Fourier Transform of the harmonic field along the azimuthal coordinate, and we integrate its modulus squared over the radial coordinate. In Fig. 3 we plot the OAM content of the LCP (red) and RCP (blue) harmonic emission in SLG driven by (a) a $\mathcal{P} = 1$ and (b) a $\mathcal{P} = 2$ LPVB as a function of the harmonic order. For the sake of comparison we include in Fig. 3c results in Ar driven by a $\mathcal{P} = 1$ LPVB. In this later case each polarization mode of the harmonic field is composed by the same OAM components as in the driving field [41]. It is worth to mention that, for the same gas target driven by a vector-vortex beam, i.e. a vector beam with non-zero net topological charge, it has been shown that the harmonic topological charge scales linearly with the OAM charge of the driver [42, 52]. The comparison of the results in Fig. 3 demonstrates that the harmonic build-up in graphene is far more complex than in isotropic targets. The OAM content of each polarization mode is extended in steps $\Delta \ell = \pm 6\ell$, a consequence of graphene's 6-fold rotational symmetry. Note also that all harmonic orders present the same OAM content.

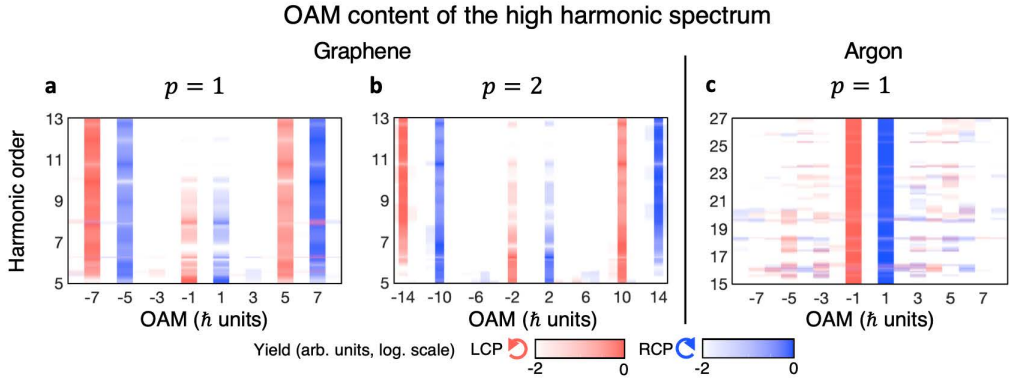

FIG. 3. OAM content of the LCP and RCP components of the high harmonic spectrum driven by a LPVB with $\mathcal{P} = 1$ (a) and with $\mathcal{P} = 2$ (b) in graphene. The interplay between the driving beam and crystal symmetries leads to the appearance of higher components—different from those of the driver—in the OAM content of the harmonic beams. For comparison, the OAM content obtained in argon driven by a LPVB with $\mathcal{P} = 1$ is shown in (c). The OAM is extracted from the Fourier Transform of the harmonic field along the azimuthal coordinate.

## III. DISCUSSION

In order to understand the far-field characteristics of the harmonic emission presented in the previous section, we derive an analytical model by means of a Fraunhofer integration that demonstrates the coupling between the target symmetries and the driving field's topology. Such understanding allows us to propose a topological harmonic spectroscopy scheme by solving the *inverse problem*: to identify the crystal's nonlinear response properties through the topological structure of the far-field harmonic emission. This method is derived to consider any crystalline structure, though we have validated it in the case of SLG.

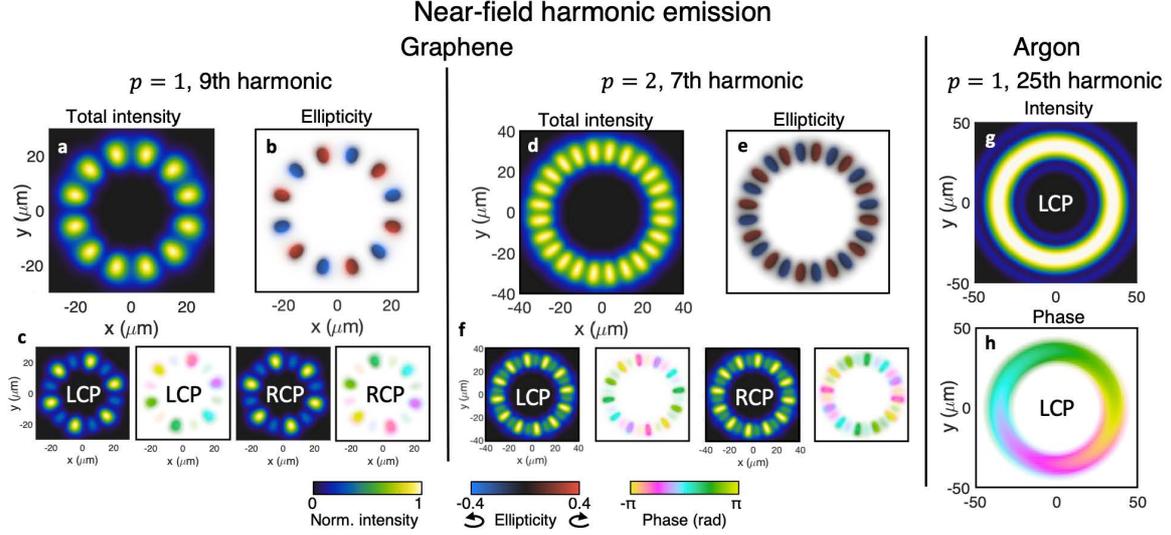

FIG. 4. Near-field intensity, ellipticity and phase properties of the 9th harmonic (a-c) and 7th harmonic (d-f) of SLG driven by a $\mathcal{P}=1$ and $\mathcal{P}=2$ LPVB, respectively. The decomposition into polarization components evidences the connection through a mirror transformation of the LCP and RCP intensity and phase distributions. For the sake of comparison, (g, h) show the near-field LCP component's intensity and phase distributions of the 25th harmonic in argon driven by a $\mathcal{P}=1$ LPVB. The ellipticity $\epsilon$ is computed from the Stokes parameters, where $\epsilon = \pm 1$ corresponds to LCP (+) and RCP (-) fields, respectively.

## A. Understanding the coupling between the target symmetries and the driving fields topology

In this section we derive an analytical model for the high harmonic far-field profile that allows us to obtain and understand its properties from the near-field harmonic emission. We show in Fig. 4 the near-field intensity and ellipticity distributions of the 9th and 7th harmonics obtained from our simulations of SLG, driven by $\mathcal{P}=1$ and $\mathcal{P}=2$ LPVBs, respectively, as well as the intensity and phase profiles of the RCP and LCP components. The amplitudes of both polarization components are connected by a mirror transformation, $\varphi \to -\varphi$. This symmetry is inherited from the driving field, and it is preserved during HHG, as a consequence of the mirror reflection symmetry stemming from the point group of graphene, namely the dihedral group $\mathbf{D}_{6h}$.

For the sake of comparison we show in Fig. 4c the intensity and phase profiles of the LCP component of the 25th harmonic obtained in Ar driven by a $\mathcal{P} = 1$ LPVB. In this case, as for any isotropic target, the RCP component shows identical intensity profile as the LCP, and conjugated phase. As a consequence, the superposition of both polarization modes results in an LPVB harmonic near-field with the same topology, same $\mathcal{P}$, as the driver. Therefore, HHG in gases can be regarded as a topologically invariant frequency up-conversion of the driving field—$\mathcal{P}$ being the topological invariant—resulting from the isotropic character of the non-linear response of the gas.

In sharp contrast, the harmonic near-field intensity obtained from SLG is structured into a necklace pattern (Figs. 4a and 4d). The necklace beads correspond to target regions where the driver's polarization tilt coincides with those angles where the SLG anisotropy shows a stronger non-linear response [47]. In correspondence, the harmonic phase and ellipticity distributions are modulated, as shown in Figs. 4(a-f). Note therefore that, as a result of the SLG anisotropic non-linear response, the harmonic near-field emitted by SLG does not correspond anymore to a LPVB, meaning that the $\mathcal{P}$ topological invariance in HHG is broken.

According to Fig. 3, the amplitude of each polarization mode of the harmonic field can be cast into a superposition of OAM modes. Thus, in the general case of a target of $N$-fold symmetry, the near field amplitude can be expressed as

$$F_q^\pm(\rho, \varphi) = e^{\pm i\ell\varphi} \sum_{s=-\infty}^{\infty} c_{q,s}^\pm(\rho) \, e^{\pm iNs\ell\varphi}, \qquad (2)$$

where $q$ is the harmonic order, and $c_{q,s}^\pm(\rho)$ are complex Fourier amplitudes. The upper/lower sign in Eq. (2) applies to the RCP/LCP components of the field, respectively. Factoring the near-field as a product of the driving field amplitude times the material response function (susceptibility), —i.e. the first factor and the sum term in Eq. (2)— the susceptibility $\chi_q$ of the non-linear response of the target is defined by the near-field amplitudes as

$$\chi_q^{\pm}(\rho,\varphi) \propto \sum_{s=-\infty}^{\infty} c_{q,s}^{\pm}(\rho) e^{\pm iNs\ell\varphi} \quad (3)$$

Interestingly, the rotational symmetry of the target response reflects the coupling between the target's symmetry, $\mathbf{C}_N$, and the driving field Poincaré's topological index, $\mathcal{P} = \ell$, thus revealing the fundamental coupling between the target symmetries and the driving field's topology in HHG.

Taking into account the near-field harmonic description, we now derive an analytical model to reproduce the far-field harmonic profile, in order to understand its topological properties. To do this, we simplify the near-field harmonic emission $F_q^{\pm}(\rho,\varphi)$ in Eq. (2) to a circumference with amplitude $F_q^{\pm}(\rho_0,\varphi)$, $\rho_0$ corresponding to the radius of maximum intensity in Figs. 4a and 4d. We use this expression to compute the Fraunhofer integral for the far-field distribution (see Methods IV B), which reads as

$$U_q^{\pm}(\beta,\phi) = \sum_s \eta_{q,s}^{\pm} J_{\pm(Ns+1)\ell}(\kappa\beta) e^{\pm i(Ns+1)\ell\phi} \quad (4)$$

where $(\beta,\phi)$ are the radial and azimuthal far-field angular coordinates, and $\kappa = 2\pi q\rho_0/\lambda$, $\lambda$ being the driver's wavelength. The coefficients $\eta_{q,s}^{\pm}$ are complex amplitude factors proportional to the near-field Fourier components $c_{q,s}^{\pm}(\rho_0)$: $\eta_{q,s}^{\pm} = -2\pi i \frac{q\rho_0}{\lambda D} e^{i2\pi qD/\lambda} c_{q,s}^{\pm}(\rho_0) e^{\mp i(Ns+1)\ell\pi/2}$. In Fig. 5a we plot the far-field intensity profile of the LCP component of the 9th harmonic from SLG computed from our model—Eq. (4) using $N = 6$. The excellent agreement between the main features of the results from our simplified model and the the exact results (Fig. 2a) allows us to use this model to analyze the topological structure of the far-field harmonics.

### B. Topological harmonic spectroscopy

Inspired by the results presented in Figs. 2 and 5a, we propose to decompose a general far-field harmonic profile into a topological cluster. Indeed, the far-field profile in Eq. (4) can be rewritten as the superposition of vortices. Such cluster is composed by the repetition of a single elemental vortex structure, with topological charge $\ell$ and radius $a_0$. First, the central component of the cluster propagates on axis, therefore it is given by

$$U_q^{\pm,0}(r,\phi) = A_0 J_{\pm\ell}\left(z_\ell \frac{r}{a_0}\right) e^{\pm i\ell\phi}, \quad (5)$$

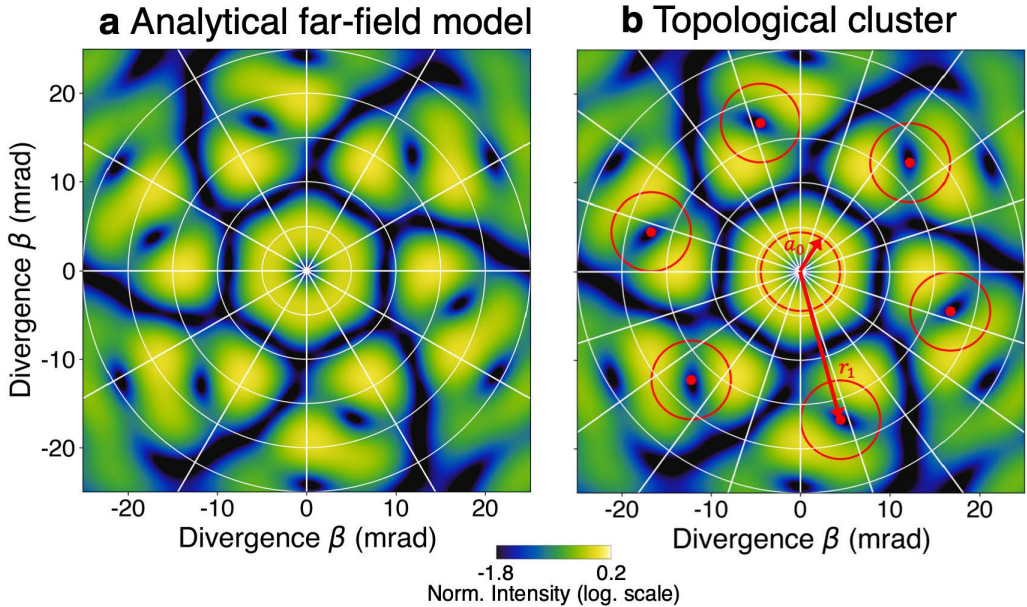

FIG. 5. Harmonic far-field from our simplified model and its reconstruction using a topological cluster of vortices. **a** Intensity profile of the LCP component of the 9th harmonic from SLG driven by a $\mathcal{P} = 1$ LPVB, obtained from the analytical far-field model, given by Eq. (4). The resulting far-field profile can be decomposed into a topological cluster of vortices with $\ell = 1$ and radius $a_0$, as depicted by the red lines in panel **b**: a central vortex and a necklace of radius $r_1 = 4.1 a_0$ composed of $N\ell = 6$ vortices. The excellent agreement of the resulting far-field intensity profile reconstructed from the topological cluster in panel **b**, given by Eq. (8), compared to panel **a** and Fig. 2a, demonstrates the working principle of topological harmonic spectroscopy.

where $z_\ell$ is the position of the first amplitude maximum of the Bessel function $J_\ell(z)$, and $r$ and $\phi$ are the polar coordinates of the far-field plane. Note that the far-field divergence $\beta$ and the radial coordinate $r$ are related through the distance to the detector, $D$, by $\beta \approx r/D$. Second, the other vortices composing the cluster, present diverging centers and are organized as set of necklaces with radii $r_\nu$, $\nu > 0$ being the necklace index (see red lines in Fig. 5a). Each necklace $\nu$ is composed by a regular distribution of $N\ell$ vortices, placed at azimuhal angles $\phi_{n,\nu}^\pm = 2\pi n/N\ell \pm \phi_{0,\nu}$. Therefore, the $\nu$ necklace field is given by

$$U_q^{\pm,\nu}(r,\phi) = A_\nu \sum_{n=0}^{N\ell-1} J_{\pm\ell}\left(z_\ell \frac{\sqrt{(x-x_n^{\pm,\nu})^2 + (y-y_n^{\pm,\nu})^2}}{a_0}\right) e^{\pm i\ell \arctan \frac{y-y_n^{\pm,\nu}}{x-x_n^{\pm,\nu}}} \quad (6)$$

where $x = r\cos\phi$, $y = r\sin\phi$ are the far-field cartesian coordinates, and $x_n^{\pm,\nu} = r_\nu \cos\phi_n^\pm$ and $y_n^{\pm,\nu} = r_\nu \sin\phi_n^\pm$ denote the position of the vortex centers within the necklace.

Fig. 5b shows the resulting superposition of the central vortex and the first necklace, with the choices $A_1/A_0 = 0.55 e^{-i0.45\pi}$, $r_1 = 4.1 a_0$, and $\phi_{0,1} = 12°$. As it can be observed, an excellent description of the far-field harmonic profile can be given equivalently either by a polar distribution of a set of on-axis vortices—obtained through the analytical far-field model given by Eq. (4)—or by a topological cluster composed of identical vortices distributed as necklaces around a central one—given by Eq. (6). Note that from the practical viewpoint, this later representation is described by geometrical parameters: the vortex and necklace radii ($a_0$ and $r_\nu$), and the necklace rotation ($\phi_{0,\nu}$), which can be well determined by a simple inspection of the far-field intensity profile. On the other hand, the contrast ratios $A_\nu/A_0$ can be also found by best fit from the far-field intensity distribution.

The configuration of the far-field topological cluster, therefore, encodes the details of the non-linear response of the target, as depicted by Eq. (3), defining topology as a relevant spectroscopic observable. The suitability of the topological approach can be demonstrated by determining the direct relationship between the geometrical parameters of the far field vortices and the Fourier components ($c_s$) describing the material response (see Eq. (3)). To this aim, we use the equivalence between the far-field descriptions: the polar description of on-axis vortices—Eq. (4)—and the topological cluster composed of necklaces of displaced vortices—Eq. (6). As we demonstrate in the Methods section, the necklace of displaced vortices $\nu$ can be re-written as

$$U_q^{\pm,\nu}(r,\phi) = A_\nu N\ell e^{\pm i\ell\phi} \sum_s e^{\pm i N s\ell(\phi \mp \phi_{0,\nu})} J_{\pm Ns\ell}\left(z_\ell \frac{r_\nu}{a_0}\right) J_{\pm(Ns+1)\ell}\left(z_\ell \frac{r}{a_0}\right). \qquad (7)$$

Considering the on-axis vortex, which is given by $U_q^{\pm,0}(r,\phi) = A_0 J_{\pm\ell}\left(z_\ell \frac{r}{a_0}\right) e^{\pm i\ell\phi}$, the total far field can be expressed as

$$\begin{aligned} U_q^\pm(r,\phi) &= \sum_\nu U_q^{\pm,\nu}(r,\phi) \\ &= N\ell \sum_s \left[\sum_{\nu>0} A_\nu e^{-iNs\ell\phi_{0,\nu}} J_{\pm Ns\ell}\left(z_\ell \frac{r_\nu}{a_0}\right)\right] J_{\pm(Ns+1)\ell}\left(z_\ell \frac{r}{a_0}\right) e^{\pm i(Ns+1)\ell\phi} \\ &\quad + A_0 J_{\pm\ell}\left(z_\ell \frac{r}{a_0}\right) e^{\pm i\ell\phi}. \end{aligned} \qquad (8)$$

Comparing Eq. (8) to Eq. (4), we find two relevant relationships. On the one hand, through inspection of the arguments of the Bessel functions of order $\pm(Ns+1)\ell$ we can extract the ratio between the radius of the elemental vortex composing the far-field topological cluster ($a_0$) and the distance from the target to the far-field plane ($D$) as

$$\frac{a_0}{D} = \frac{z_\ell \lambda}{2\pi q \rho_0}. \tag{9}$$

This allows us to establish a direct relationship between the vortex radii, $a_0$, with the radius of the driving LPVB, $\rho_0$. On the other hand, a second condition is given by

$$c_{q,s}^{\pm}(\rho_0) = i^{\pm Ns\ell} K_\ell^{\pm} N\ell \sum_{\nu>0} A_\nu e^{-iNs\ell\phi_{0,\nu}} J_{\pm Ns\ell}\left(z_\ell \frac{r_\nu}{a_0}\right) \text{ for } s \neq 0 \tag{10}$$

$$c_{q,0}^{\pm}(\rho_0) = K_\ell^{\pm}\left[A_0 + N\ell \sum_{\nu>0} A_\nu J_0\left(z_\ell \frac{r_\nu}{a_0}\right)\right], \tag{11}$$

with $K_\ell^{\pm} = i^{\pm\ell+1} e^{-i2\pi qD/\lambda} a_0/z_\ell$. Thus, Eqs. (10) and (11) demonstrate that the target response, Eq. (3), is completely defined by the characteristics of the topological objects that describe the harmonic far field. Additionally, Eqs. (10) and (11) ground a basic procedure for topological spectroscopy: once the topological structure of the harmonic far-field is recorded, the nonlinear response of the target can be extracted. In particular, by measuring the number ($\nu$) and rotation ($\phi_{0,\nu}$) of the necklaces, the necklace to vortex radii ratio ($r_\nu/a_0$), and the amplitudes ratio of the necklace vortices to the central one ($A_\nu/A_0$), and assuming vortices with Bessel profiles—as those shown in Eq. (6)—, Eqs. (10) and (11) can be used to recover the Fourier components of the target response in a circle of radius $\rho_0$—i.e. the circle of intensity maxima of the driving LPVB. The direct map of the driver's azimuth to the polarization, characteristic in a LPVB, allows using the inverse Fourier transform of the coefficients $c_{q,s}^{\mp}$ to recover the q-th harmonic non-linear anisotropic response of the target. In addition, it holds the potential to uncover the possible inhomogeneous response of the target response along the circle of maximum intensity of the driver.

As a proof of concept, we apply the above steps to the 9th order harmonic far field obtained numerically from an $\mathcal{P} = 1$ LPVB driver with $\rho_0 = 22$ $\mu$m, as shown in Figs. 2a and 2b. As we mentioned above, simple best-fit analysis yields a far-field cluster composed by an on-axis vortex and a single necklace of $N\ell = 6$ vortices, with rotation $\phi_{0,1} = 12°$, an amplitude ratio $A_1/A_0 = 0.55 e^{-i0.45\pi}$, and a radius $r_1 = 4.1 a_0$. For these parameters, Eq. (9) gives $a_0/D = 4.44$ mrad. Feeding Eqs.(10) and (11) with these necklace and vortex parameters we can find the values of the 9th-harmonic response coefficients $c_{9,0}, c_{9,\pm 1}$ defined in Eq. (3). The relative ratios found are $c_{9,1}/c_{9,0} = \pm 0.84 \times e^{i0.22\pi}$ and $c_{9,-1}/c_{9,0} =$ $\pm 0.83 \times e^{-i0.87\pi}$. Taking into account that the ratios between the Fourier components of the computed near-field shown in Fig. 5 are $c_{9,1}/c_{9,0} = 0.83 \times e^{i0.35\pi}$ and $c_{9,-1}/c_{9,0} = 0.83 \times e^{-0.84\pi}$, and that we have only considered the central vortex, $\nu = 0$ and the first necklace $\nu = 1$, our results demonstrate the potentiality of topological high harmonic spectroscopy to extract information about the anisotropic crystal's response. In the Supplementary Information we demonstrate that this method can be applied to identify the role of interband and intraband contributions to HHG. However, we note that the proposed method would highly benefit from further developments on retrieval algorithms that can infer the anisotropic response through topological far-field traces.

### C. Conclusion

We have demonstrated a new scenario for high-harmonic spectroscopy stemming from the interaction of structured driving beams with crystalline solid targets. In contrast to isotropic gaseous targets, we show that crystal symmetries couple with the driving beam's topology during HHG. The signature of this coupling is encoded into a complex spatial structure in the emitted harmonics. Particularly, we unveil this intertwined photon conversion by studying HHG from single-layer graphene driven by LPVB. We show that, in contrast to the isotropic case, the harmonics generated from crystal targets can break the conservation of the driver's topological structure, according to their constituent symmetries. We provide for an analytical derivation that allows to (i) predict the topology of the high harmonic beams from the targets anisotropic symmetry, and (ii), to retrieve the anisotropic response of the target from the topology of the high harmonic beams. As a consequence, high harmonic spectroscopy based on topology allows to extract *spatially* resolved information about the nonlinear response of the target, which can not be obtained with standard spectroscopic techniques.

Though we have demonstrated the interplay of the vector beam driver topology with the target's symmetries in two-dimensional materials such as graphene, we believe our results open a new general scenario for topological optics in which the target's non-linear response is coupled with the topological structure of light. In this sense the scenario of HHG in bulk crystals [44, 45] is of particular interest, as propagation effects may play a relevant role. In general, any property that presents an anisotropic HHG response could be characterized. For example, interband and intraband contributions to HHG respond differently to the drivers ellipticity in bulk silicon [44], and as such the nature of the harmonics contribution can be characterized through its topology when driven by properly chosen vector beams (see Supplementary information). Finally, we believe that this technique can be further used to characterize more complex targets such as polycrystals [53, 54] or heterostructures [55].

## IV. MATERIALS AND METHODS

### A. Numerical simulations of high-harmonic generation in graphene and in gases.

HHG driven by structured beams requires the computation of the macroscopic response of the target. Our strategy follows the discrete-dipole approximation method presented in [56], that has been recently also applied to graphene polycrystals [57]. In this method, the graphene target is divided into a set of elemental surfaces of dimensions small enough to assume the local field profile constant, but still enclosing a sufficient number of graphene's lattice cells to allow the approximation of the Brillouin zone as a continuous region. Next, we integrate the Schrödinger equation to obtain the mean dipole acceleration in each elementary surface. The dipole acceleration is used to compute the time derivative of the current density, which is proportional to the radiated near-field, and used as a source for the electromagnetic field propagator, in order to find the far-field distribution. Dynamics of the SLG interaction with the driving field is integrated from the Schrödinger equation in the nearest neighbor tight-binding approximation [12, 56]. We have also implemented the laser-driven dynamics in SLG through the semiconductor Bloch equations (SBE). The comparison between the macroscopic TDSE and SBE simulations (see Supplementary Information) demonstrates that our results do not depend on the formalism used to calculate the current density.

In order to compare the results of HHG from SLG with that from an isotropic target, we have conducted calculations of HHG in an infinitely thin Ar gas jet. For this, we have followed the method presented in Ref. [58], that has been successfully validated against several experiments (see for example [24, 25, 27, 36–38, 41, 52]). Similarly to the procedure used for SLG, discussed in the above paragraph, the gas target is split into elemental emitters. The dipole acceleration in each emitter is computed using the strong field approximation, without resorting to the saddle-point approximation.

### B. Derivation of the analytical far-field model

In our model we consider a simplified representation of the $q$-th harmonic near-field as a circumference of radius of maximal amplitude $\rho_0$, which also corresponds to the radius of maximum intensity of the the driving field, as

$$R_q^\pm(\rho,\varphi) = F_q^\pm(\rho,\varphi)\delta(\rho - \rho_0), \tag{12}$$

with $F_q^\pm(\rho,\varphi)$ the near-field azimuthal profile, Eq. (2), and $(\rho,\varphi)$ the near-field radial and azimuthal coordinates, respectively. In the results presented in this work for SLG, $\rho_0$ corresponds to 22 $\mu$m for the $\mathcal{P} = 1$ LPVB, and 25 $\mu$m for the $\mathcal{P} = 2$ LPVB.

We compute the harmonic far-field amplitude, $U_q^\pm(\beta,\phi)$, from the Fraunhofer integral of the near-field $R_q^\pm(\rho,\varphi)$. Using Eq. (2), and after the trivial radial integration, we obtain

$$U_q^\pm(\beta,\phi) = -iq\frac{e^{i2\pi Dq/\lambda}}{\lambda D}\rho_0 \sum_{s=-\infty}^{\infty} c_{q,s}^\pm(\rho_0) \int_0^{2\pi} d\varphi\ e^{\pm i(Ns+1)\ell\varphi}e^{-i\kappa\beta\cos(\phi-\varphi)} \tag{13}$$

with $\kappa = 2\pi q\rho_0/\lambda$, $\lambda$ being the driver's wavelength, and $(\beta,\phi)$ the far-field divergence and azimuthal angles, respectively. Using the identity

$$e^{i\alpha\cos(\phi-\varphi)} = \sum_m i^m J_m(\alpha)e^{im(\phi-\varphi)}, \tag{14}$$

the azimuthal integral in Eq. (13) leads to the condition $m = \mp(Ns+1)\ell$, and thus Eq. (13) leads to Eq. (4).

### C. Derivation of the topological cluster

A vortex necklace is a distribution of identical vortices with centers equally distributed along a ring of radius $r_\nu$. To compute the polar form of such structure, we shall first find the polar expression of an off-axis vortex with center at an arbitrary coordinate $(x_n, y_n)$, as

$$V_\ell^{(x_n,y_n)}(x,y) = V_0 J_\ell\left(\frac{z_\ell}{a_0}\sqrt{(x-x_n)^2 + (y-y_n)^2}\right) e^{i\ell\arctan\frac{y-y_n}{x-x_n}}, \tag{15}$$

where $(x,y)$ are the far-field cartesian coordinates. Eq. (15) corresponds to the translation of the on-axis vortex $V_\ell^{(0,0)}$ to the point $(x_n, y_n)$. To compute the polar form of Eq. (15), we apply the translation as a phase shift in Fourier space,

$$\tilde{V}_\ell^{(x_n,y_n)}(k_x, k_y) = e^{-ik_x x_n}e^{-ik_y y_n}\tilde{V}_\ell^{(0,0)}(k_x, k_y). \tag{16}$$

Defining the far-field polar coordinates in real and Fourier spaces as $(r, \phi)$ and $(k, \varphi)$, respectively, we use the coordinate transformations $x = r \cos \phi$, $y = r \sin \phi$, $k_x = k \cos \varphi$ and $k_y = k \sin \varphi$, to compute $\tilde{V}_\ell^{(0,0)}$ in polar coordinates as

$$\tilde{V}_\ell^{(0,0)}(k, \varphi) = \frac{V_0}{2\pi} \iint J_\ell \left( z_\ell \frac{r}{a_0} \right) e^{i\ell\phi} e^{-ikr\cos(\phi-\varphi)} r dr d\phi$$
$$= i^{-\ell} V_0 e^{i\ell\varphi} \int J_\ell \left( z_\ell \frac{r}{a_0} \right) J_\ell(kr) r dr = i^{-\ell} V_0 \frac{a_0}{z_\ell} \delta \left( \frac{z_\ell}{a_0} - k \right) e^{i\ell\varphi} \quad (17)$$

where, for the last step, we have used the identity

$$\frac{1}{x} \delta(x - a) = \int_0^\infty J_\ell(xt) J_\ell(at) t dt. \quad (18)$$

Defining $x_n = r_n \cos \phi_n$ and $y_n = r_n \sin \phi_n$, we can compute $V_\ell^{(x_n, y_n)}$ in Eq. (15) in polar coordinates as the inverse Fourier transform of $\tilde{V}_\ell^{(x_n, y_n)}$ in Eq. (16), leading to

$$V_\ell^{(r_n, \phi_n)}(r, \phi) = \frac{1}{2\pi} \iint \tilde{V}_\ell^{(0,0)}(k, \varphi) e^{-ikr_n \cos(\varphi - \phi_n)} e^{ikr \cos(\varphi - \phi)} k dk d\varphi. \quad (19)$$

Using Eqs.(17) and (14), we find from Eq. (19) the polar description for the vortex displaced to $(x_n, y_n)$ as

$$V_\ell^{(r_n, \phi_n)}(r, \phi) = (-1)^\ell V_0 e^{i\ell\phi} \sum_m e^{-im(\phi-\phi_n)} J_m \left( \frac{z_\ell}{a_0} r_n \right) J_{m-\ell} \left( \frac{z_\ell}{a_0} r \right). \quad (20)$$

Correspondingly, using Eq. (20) the polar expression of the vortex necklace with radius $r_\nu$ defined in Eq. (6) is given by

$$U_q^{\pm,\nu}(r, \phi) = \sum_{n=0}^{N\ell-1} V_{\pm\ell}^{(r_\nu, 2\pi\ell \frac{n}{N} \mp \phi_{0,\nu})}(r, \phi)$$
$$= A_\nu N \ell e^{\pm i\ell\phi} \sum_s e^{\pm iNs\ell(\phi \mp \phi_{0,\nu})} J_{\pm Ns\ell} \left( \frac{z_\ell}{a_0} r_\nu \right) J_{\pm(Ns+1)\ell} \left( \frac{z_\ell}{a_0} r \right), \quad (21)$$

**ACKNOWLEDGMENTS**

We acknowledge economic support from the Spanish Ministerio de Ciencia, Innovación y Universidades (PID2019-106910GB-100). This project has also received funding from the European Research Council (ERC) under the European Unions Horizon 2020 research and innovation program (Grant Agreement No. 851201). A.G.-C. acknowledges support from Ministerio de Educación, Cultura y Deporte (FPU18/03348). C.H.-G. acknowledges Ministerio de Ciencia, Innovación, y Universidades for a Ramón y Cajal contract (RYC-2017-22745), co-funded by the European Social Fund.

**DECLARATIONS**

The authors declare no conflict of interest.

**DATA AVAILABILITY**

Data underlying the results presented in this paper are available from the authors upon reasonable request.

**CODE AVAILABILITY**

The codes used for simulations and data analysis are available from the corresponding author upon reasonable request.

---

# Supplementary information for "Interplay of crystal symmetries and light's topology in high harmonic spectroscopy"

Ana García-Cabrera,[*] Roberto Boyero-García, Óscar Zurrón-Cifuentes,
Javier Serrano, Julio San Román, Luis Plaja, and Carlos Hernández-García

*Grupo de Investigación en Aplicaciones del Láser y Fotónica,*
*Departamento de Física Aplicada, Universidad de Salamanca, E-37008, Salamanca, Spain*

(Dated: November 3, 2023)

## Abstract

In this Supplementary Information we include: (i) simulations of HHG in single-layer graphene using the two-band semiconductor Bloch equations approach, including both the microscopic and macroscopic points of view; (ii) simulations that show the role of interband and intraband contributions in high harmonic spectroscopy based on topology; and (iii) simulation results of HHG in single-layer graphene driven by vector beams for other harmonic orders than those shown in the main text.

[*] anagarciacabrera@usal.es

# I. SEMICONDUCTOR BLOCH EQUATIONS

To demonstrate the role of decoherence in our results, and to show the accuracy of our TDSE-based method, we have implemented the semiconductor Bloch equations (SBE) into our macroscopic approach. The equations for the two-level density matrix elements, in which we include the dephasing-time term $T_2$, read as:

$$i\hbar\dot{\rho}_{++}(\boldsymbol{\kappa_t}, t) = \boldsymbol{F}(t)\boldsymbol{D}(\boldsymbol{\kappa_t})\left[\rho_{-+}(\boldsymbol{\kappa_t}, t) - \rho_{+-}(\boldsymbol{\kappa_t}, t)\right], \tag{1}$$

$$i\hbar\dot{\rho}_{--}(\boldsymbol{\kappa_t}, t) = \boldsymbol{F}(t)\boldsymbol{D}(\boldsymbol{\kappa_t})\left[\rho_{+-}(\boldsymbol{\kappa_t}, t) - \rho_{-+}(\boldsymbol{\kappa_t}, t)\right], \tag{2}$$

$$i\hbar\dot{\rho}_{+-}(\boldsymbol{\kappa_t}, t) = \left[E_{-}(\boldsymbol{\kappa_t}) - E_{+}(\boldsymbol{\kappa_t}) - i\frac{\hbar}{T_2}\right]\rho_{+-}(\boldsymbol{\kappa_t})$$
$$+ \boldsymbol{F}(t)\boldsymbol{D}(\boldsymbol{\kappa_t})\left[\rho_{--}(\boldsymbol{\kappa_t}, t) - \rho_{++}(\boldsymbol{\kappa_t}, t)\right], \tag{3}$$

$$i\hbar\dot{\rho}_{-+}(\boldsymbol{\kappa_t}, t) = \left[\rho_{+-}(\boldsymbol{\kappa_t}, t)\right]^{*}, \tag{4}$$

where $\rho_{i,j}$ are the density matrix elements, $\boldsymbol{\kappa_t}$ is the quasimomentum $\boldsymbol{k}$ transformed to a frame that moves with the vector potential, $\boldsymbol{\kappa_t} = \boldsymbol{k} - \frac{q_e}{\hbar c}\boldsymbol{A}(t)$, $\boldsymbol{F}(t)$ is the laser field amplitude, $\boldsymbol{D}(\boldsymbol{\kappa_t})$ is the transition dipole moment, $E(\boldsymbol{\kappa_t})$ is the energy and the signs $+$ and $-$ refer to the conduction and valence bands, respectively. To elude the divergence of the terms $\boldsymbol{D}(\boldsymbol{\kappa_t})$ at the Dirac points, we transform these equations to a new base, as described in [1].

Equations (1) to (3) are equivalent to the so-called SBE within the density matrix formalism given in [2]:

$$\dot{\pi}(\boldsymbol{\kappa_t}, t) = -\frac{\pi(\boldsymbol{\kappa_t}, t)}{T_2} - i\Omega(\boldsymbol{\kappa_t}, t)w(\boldsymbol{\kappa_t}, t)e^{-iS(\boldsymbol{\kappa_t}, t)}, \tag{5}$$

$$\dot{n}_m(\boldsymbol{\kappa_t}, t) = is_m\Omega^{*}(\boldsymbol{\kappa_t}, t)\pi(\boldsymbol{\kappa_t}, t)e^{iS(\boldsymbol{\kappa_t}, t)} + c.c., \tag{6}$$

where $\pi(\boldsymbol{\kappa_t}, t)$ is the term of coherence between the two levels, $n_m(\boldsymbol{\kappa_t}, t)$ is the population in the conduction or valence band, $\Omega(\boldsymbol{\kappa_t}, t)$ is the Rabi frequency, $w(\boldsymbol{\kappa_t}, t)$ is the difference between the population in the valence band and the conduction band, $S(\boldsymbol{\kappa_t}, t)$ is the classical action and $s_m$ is $-1$ for the valence band and $+1$ for the conduction band, assuming that, before the interaction, all the electrons are in the valence band.

The equivalence between the two sets of equations can be easily verified by identifying $\pi(\boldsymbol{\kappa_t}, t)$ with $\rho_{+-}(\boldsymbol{\kappa_t}, t)e^{-iS(\boldsymbol{\kappa_t}, t)/\hbar}$, the Rabi frequency $\Omega(\boldsymbol{\kappa_t}, t)$ with $\hbar\boldsymbol{F}(t)\boldsymbol{D}(\boldsymbol{\kappa_t})$ and $w(\boldsymbol{\kappa_t}, t)$ with $\rho_{--}(\boldsymbol{\kappa_t}, t) - \rho_{++}(\boldsymbol{\kappa_t}, t)$.

The dipole acceleration obtained within the SBE formalism is integrated into the macro-scopic HHG response in the same way as it was done for the TDSE (see Methods in the main text).

In Fig. S1, we show the comparison between the macroscopic TDSE and SBE HHG results in single-layer graphene driven by a radially polarized beam ($\mathcal{P} = 1$). We have considered two different values of the decoherence time, $T_2$, for the spatially-integrated far-field spectrum (a) and the far-field normalized intensity (b) and phase (c) distributions of the 9th harmonic's RCP component. Note that $T_2 = 35$ fs is the decoherence time proposed for graphene in a recent work [3], based on the experimental results of [4]. As we can see in Fig. S1, the integrated spectra resulting from the TDSE and the SBE formalisms with $T_2 = 35$ fs are in excellent agreement. More importantly, in the context of our work, the far-field distributions in panels (b) and (c) barely change with the dephasing time, even in the extreme case of $T_2 = 2.5$ fs, which confirms the validity of our results regardless of the method employed.

## II. TOPOLOGICAL CHARACTERIZATION OF INTERBAND AND INTRA-BAND SUB-CYCLE DYNAMICS

As it has been introduced in the main text, HHG from solids can be interpreted in terms of semi-classical trajectories of electron-hole pairs in the crystal that evolve accordingly to the band's energy dispersion. The harmonic emission is governed by the radiation upon recombination of the electron-hole pair from the conduction and valence bands. These contributions are known as interband harmonics. However, solid systems also present contributions to the HHG spectrum from intraband dynamics [2]. The experimental distinction between interband and intraband harmonic contributions is not trivial. In this section we demonstrate that interband and intraband contributions can be distinguished through topological harmonic spectroscopy.

Topological spectroscopy is sensitive to the anisotropy. Thus, if intraband and interband contributions present different anisotropy in their HHG nonlinear response, they could be distinguished by characterizing the topology of the resulting harmonic beam. For example, it has been reported that interband and intraband mechanisms respond differently to the drivers ellipcitiy in bulk silicon, and their relative contribution also differs with the harmonic order [5]. Thus, HHG in bulk silicon driven by properly chosen vector beams would result in high-order harmonic beams whose topology is unequivocally related to their interband or intraband nature.

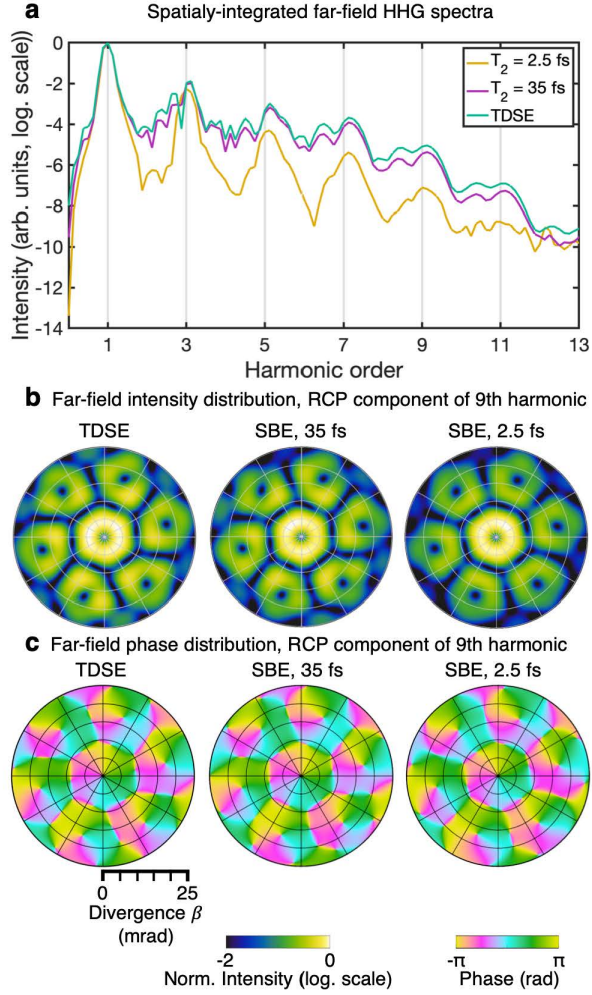

FIG. S1. **a** Spatially-integrated far-field HHG spectrum, and far-field **b** normalized intensity (logarithmic scale) and **c** phase distributions of the RCP component of the 9th harmonic generated in single-layer graphene by a radially polarized beam ($\mathcal{P} = 1$). Three simulations results are presented depending on the simulation approach: TDSE, SBE with dephasing time of 35 fs (which corresponds to the predicted dephasing time in graphene according to [3]), and SBE with dephasing time of 2.5 fs.

We have performed simulations of HHG in single-layer graphene to demonstrate how interband and intraband dynamics can be distinguished through their topology.

In Fig. S2a we show the spatially integrated HHG far-field spectrum driven by a radially polarized vector beam ($\mathcal{P} = 1$) in single-layer graphene, depicting the total (dark blue), interband (green) and intraband (light blue) contributions. The driving beam parameters are as those in the main text (28 fs temporal duration, 3 $\mu$m wavelength, $5 \times 10^{10}$ W/cm$^2$ peak intensity and 30 $\mu$m beam waist). We note that the high-order harmonics are dominated by the interband dynamics, and only in the 3rd harmonic, the intraband contribution is similar to that of the interband. Thus, in Fig. S2b we show the anisotropic response of the interband and intraband contributions for the 3rd and 9th harmonic orders. To do so, we perform microscopic HHG calculations for a linearly polarized laser field, varying the tilt-angle. Whereas the behaviour of the 3rd harmonic is mainly isotropic for both contributions, interband and intraband dynamics present a different anisotropic behavior in the 9th harmonic order.

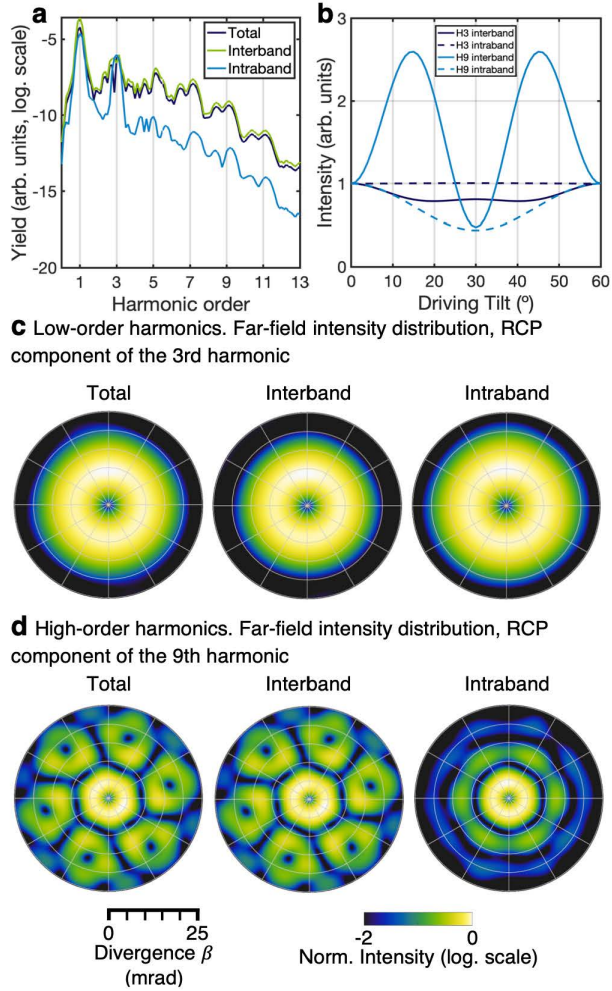

FIG. S2. **a** Spatially-integrated far-field HHG spectrum, **b** intensities of the 3rd and 9th harmonics as a function of the driver's polarization tilt (for each harmonic, intensities are normalized to the value at 0°), and far-field intensity distributions of the RCP component of the **c** 3rd and **d** 9th harmonic orders generated in single-layer graphene by a radially polarized beam ($\mathcal{P} = 1$). The role of intraband and interband contributions is explicitly shown.

Figs. S2c and S2d present the far-field intensity distribution of the RCP component of the 3rd and 9th harmonics, respectively, when driven by a radially polarized ($\mathcal{P} = 1$) laser beam. In each figure, we show the interband and intraband contributions, and the total harmonic beam. Whereas in the 3rd harmonic beam interband and intraband contributions can not be distinguished as both present similar isotorpic behaviour, in the 9th harmonic they can be clearly distinguished. Indeed, we can easily identify that the harmonic beam is dominated by the interband contributions.

### III. RESULTS FOR DIFFERENT HARMONIC ORDERS

In the main text we show as a proof of concept the HHG results in graphene for a certain harmonic order driven by LPVB with $\mathcal{P} = 1$ and $\mathcal{P} = 2$. For completeness, we show here the results for all harmonic within the HHG spectrum. In Fig. S3 we show the far-field intensity and phase distributions for the 5th, 7th, 9th and 11th harmonic orders, from top to bottom, driven by an LPVB with $\mathcal{P} = 1$ (a and b) and $\mathcal{P} = 2$ (c and d). Although the crystal symmetries are encoded in all the harmonics presenting anisotropic behaviour (5th order and above), the specific far-field spatial distribution depends on the details of the anisotropic response for each harmonic.

---

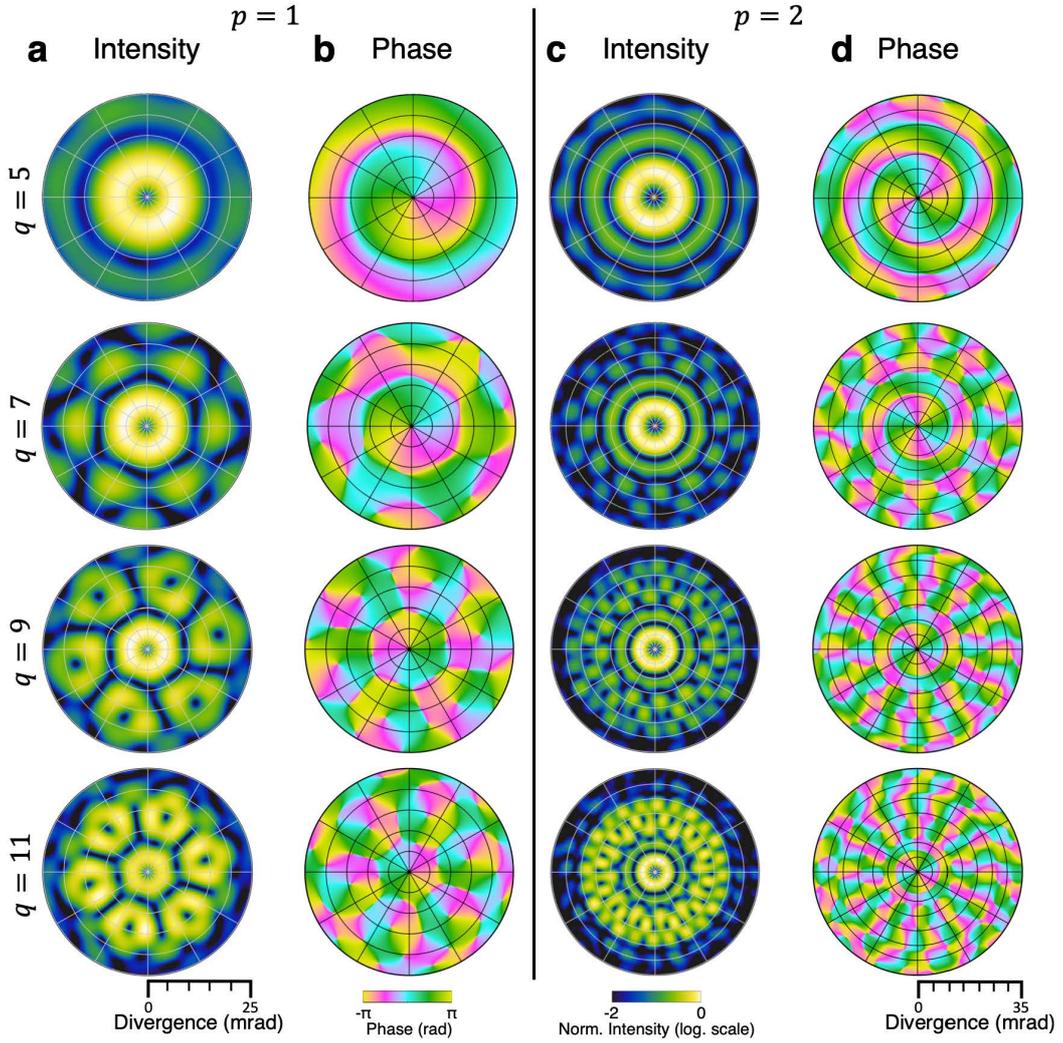

FIG. S3. Far-field intensity (**a** and **c**) and phase (**b** and **d**) distributions for the RCP component of the 5th, 7th, 9th and 11th harmonic orders obtained in graphene, driven by a vector beam with $\mathcal{P} = 1$ (**a** and **b**) and a vector beam with $\mathcal{P} = 2$ (**c** and **d**.)

7